\documentclass[twocolumn,prl,aps,superscriptaddress,showpacs]{revtex4}
\usepackage{graphicx}
\begin{document}

\title{Explicit demonstration of nonabelian anyon, braiding matrix and fusion
rules in the Kitaev-type spin honeycomb lattice models}

\author{Yue Yu}
\author{Tieyan Si}
\affiliation{Institute of Theoretical Physics, Chinese Academy of
Sciences, P.O. Box 2735, Beijing 100190, China}
\date{\today}
\begin{abstract}
The exact solubility of the Kitaev-type spin honeycomb lattice
model was proved by means of a Majorana fermion representation or
a Jordan-Wigner transformation while the explicit form of the
anyon in terms of Pauli matrices became not transparent. The
nonabelian statistics of anyons and the fusion rules can only be
expressed in indirect ways to Pauli matrices. We convert the
ground state and anyonic excitations back to the forms of Pauli
matrices and explicitly demonstrate the nonabelian anyonic
statistics as well as the fusion rules. These results may instruct
the experimental realization of the nonabelian anyons. We suggest
a proof-in-principle experiment to verify the existence of the
nonabelian anyons in nature.
\end{abstract}


\maketitle

\noindent{\it Introduction } The Kitaev-type spin honeycomb
lattice models have attracted many research interests for the
possible nonabelian anyonic excitations in these exactly soluble
two-dimensional models \cite{kitaev}. There are two topological
phases for these kinds of models. The topologically trivial A
phase is  an abelian anyon phase which is equivalent to that in
Kitaev's toric code model \cite{kitaev1}. The topologically
nontrivial B phase is within the same universality class of the
Moore-Read Pfaffian states in the fractional quantum Hall state
\cite{mr,yu} and the vortex excitations are nonabelian anyons
\cite{ms,w,mr}.

In solving these kinds of models, a key technique is the usage of
the Majorana fermion representation of the spin-1/2 operators,
either via Kitaev's Majorana fermions or the Jordan-Winger
transformation \cite{fzx,ch,cn}. However, the shortcoming to
introduce these Majorana fermions is that the ground state and the
elementary excitations are hard to be expressed by the original
spin operators, i. e., Pauli matrices. Then the nonabelian fusion
rules and statistics may not be directly shown in Pauli matrices'
language \cite{ville}. Meanwhile, experimentally exciting,
manipulating and detecting anyons may be more practical by using
the spin operators, as recently suggested or done for the toric
code model \cite{han,pan,pachos,zoller,du,cirac,bloch} and for the
A phase of Kitaev honeycomb model \cite{zd,vidal}. Therefore, to
explicitly demonstrate the nonabelian anyonic statistics, one
needs to express the ground state and elementary excitations in
the spin operators. Chen and Nussinov \cite{cn} have studied a
real space form of the ground state of the Kitaev honeycomb model
and applied it to the A phase with abelian anyons. However, for
the more interesting B phase, it was not figured out yet.

In a recent work, one of us, with Wang, has provided a generalized
model of the Kitaev honeycomb model by adding three- and four-spin
couplings \cite{yu,yu1}. With this generalized Kitaev-type model,
we showed the equivalence between the B phase with the breaking of
the time reversal symmetry and the Moore-Read Pfaffian state. In
that work, we map the model in the honeycomb lattice to a spinless
fermion model in a square lattice. The task of the present paper
is mapping back the ground state and anyonic(vortex) excitations
obtained in the square lattice to their honeycomb lattice version.

After writing down these states in spin operators, i.e., Pauli
matrices, we can check the nonabelian statistics of anyons and
fusion rules of the excitations. As expected, these results agree
with those in the previous abstractive study in Kitaev's original
work \cite{kitaev}. Moreover, these explicit forms with well-known
Pauli matrices may help the readers who are not in this special
field to understand those formal descriptions made by Kitaev. It
may also instruct the experimentalists to realize these states and
verify the existence of the nonabelian anyons in nature. Similar
to the case of the abelian anyons in the toric code model
\cite{zoller, cirac, bloch}, we may also expect to excite,
manipulate and detect the nonabelian anyons in atomic spin systems
in optical lattice. We may design a proof-in-principle experiment
to show the existence of nonabelian anyon in nature by means of
the techniques proposed and  developed recently in the photon
graph state \cite{han,pan,pachos}, and a nuclear magnetic
resonance system \cite{du}. The minimal lattice needs only six
sites, which is accessible to the current experiments, as several
experimental groups have done to the toric code model
\cite{pan,pachos,du}.

\begin{figure}[htb]
\vspace{-0.5cm}
\begin{center}
\includegraphics[width=6cm]{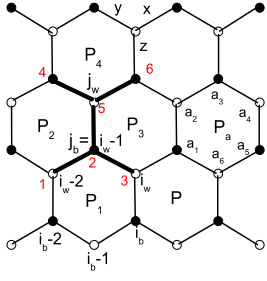}
\end{center}
\vspace{-0.6cm}
 \caption{\label{fig:Fig.1}
The honeycomb lattice and plaquette. The thick links define $A_j$
and also the proof-in-principle configurations. }
\end{figure}

\noindent{\it Model and ground state} We consider a Kitaev-type
spin model in a honeycomb lattice with a three- and four-spin
couplings \cite{yu,yu1}
\begin{eqnarray}
H&=&-J_x\sum_{x-links}\sigma_i^x
\sigma_j^x-J_y\sum_{y-links}\sigma_i^y
\sigma_j^y-J_z\sum_{z-links}\sigma_i^z \sigma_j^z\nonumber\\
&-&\kappa\sum_b
\sigma^z_b\sigma^y_{b+e_z}\sigma^x_{b+e_z+e_x}-\kappa\sum_w
\sigma^x_{w}\sigma^y_{w+e_x}\sigma^z_{w+e_x+e_z}\nonumber\\
&-&\lambda_x\sum_{b} \sigma_{b}^z\sigma^y_{b+e_z}
\sigma_{b+e_z+e_x}^y\sigma^z_{b+e_z+e_x+e_z}\nonumber\\
&-&\lambda_y\sum_{b} \sigma_{b}^z\sigma^x_{b+e_z}
\sigma_{b+e_z+e_y}^x\sigma^z_{b+e_z+e_y+e_z},
\end{eqnarray}
where  $\sigma^{x,y,z}$ are Pauli matrices, $x$-,$y$-,$z$-links
are shown in Fig.~\ref{fig:Fig.1}, $'w'$ and $'b'$ label the white
and black sites of lattice, and $e_x,e_y,e_z$ are the positive
unit vectors, which are defined as, e.g.,
$e_{a_1a_2}=e_z,e_{a_2a_3}=e_x,e_{a_6a_1}=e_y$. $J_{x,y,z}$,
$\kappa$ and $\lambda_{x,y}$ are tunable real parameters. This is
a system with the breaking of the time-reversal symmetry. There is
a $Z_2$ gauge symmetry generated by
\begin{eqnarray}
W_{P_a}=\sigma_{a_1}^x\sigma_{a_2}^y\sigma_{a_3}^z\sigma_{a_4}^x\sigma_{a_5}^y\sigma^z_{a_6}
\end{eqnarray}
where $P_a$ labels a plaquette (see Fig.~\ref{fig:Fig.1}). That
is, $[H,W_{P_a}]=0$.

We first recall how to work out the ground state of this model.
The first step is introducing the Majorana fermion representation
of the spin operators $\sigma^a$ \cite{kitaev} or employing  the
Jordan-Wigner transformation \cite{fzx,ch,cn}. For convenience, we
use the Jordan-Wigner transformation
\begin{eqnarray}
&&\psi_{i_w}=\sigma_{i_w}^y\prod_{i'_s<i_w}\sigma^z_{i'_s},~~
\psi_{i_b}=\sigma_{i_b}^x\prod_{i'_s<i_b}\sigma^z_{i'_s},\nonumber\\
&&b_{i_w}=-\sigma_{i_w}^x\prod_{i'_s<i_w}\sigma^z_{i'_s},~~
b_{i_b}=-\sigma_{i_b}^y\prod_{i'_s<i_b}\sigma^z_{i'_s}
\end{eqnarray}
where $i$ lables a $z$-link and $\psi_{i_s}$ and $b_{i_s}$ are
Majorana fermions which satisfy
$\psi_{i_s}\psi_{j_t}=-\psi_{j_t}\psi_{i_s}$,
$b_{i_s}b_{j_t}=-b_{j_t}b_{i_s}$ for $i_s\ne j_t$; and
$\psi_{i_s}^2=b_{j_t}^2=1$. $\psi_{i_s}$ and $b_{j_t}$ are also
anticommutative. The order of the sites is defined as follows:
$i_s>j_t$ if the zig-zag horizontal line that $i_s$ belongs to is
higher than that of $j_t$ or if $i_s$ is on the right hand of
$j_t$ when they are in the same line.

After this Jordan-Wigner transformation, the Hamiltonian reads
\begin{eqnarray}
H&=&-iJ_x\sum_{x-links}\psi_w\psi_b-iJ_y\sum_{y-links}\psi_b\psi_w
\nonumber\\
&-&J_z\sum_{z-links}b_wb_b\psi_b\psi_w -\kappa\sum_b
b_bb_{b+e_z}\psi_b\psi_{b+e_z+e_x}\nonumber\\&-&\kappa\sum_w
b_{w+e_x}b_{w+e_x+e_z}\psi_w\psi_{w+e_x+e_z}\\
&-&i\lambda_y\sum_b
b_bb_{b+e_z}b_{b+e_z+e_y}b_{b+e_z+e_y+e_z}\psi_b\psi_{b+e_z+e_y+e_z}\nonumber\\
&-&i\lambda_x\sum_b
b_bb_{b+e_z}b_{b+e_z+e_x}b_{b+e_z+e_x+e_z}\psi_b\psi_{b+e_z+e_x+e_z}.\nonumber
\end{eqnarray}
 According to Lieb's theorem \cite{lieb}, the ground state of the
system is $Z_2$ vortex free state. This means that
$ib_bb_{b+e_z}=1$ in the ground state sector. Following the track
in refs. \cite{cn,yu}, finally, the Hamiltonian in the ground
state sector is given by
\begin{eqnarray}
&&H_0=J_z\sum_i(d_i^\dag d_i-1/2)+
i\kappa\sum_i(d_id_{i+e_x}+d_i^\dag d_{i+e_x}^\dag)\nonumber\\&&+
\tilde J_x(d^\dag_id_{i+e_x}-d_id_{i+e_x}^\dag)+\tilde
J_y(d^\dag_id_{i+e_y}-d_id_{i+e_y}^\dag)\\&&+
\lambda_x\sum_i(d^\dag_{i+e_x}d^\dag_i-d_{i+e_x}d_i)+\tilde
\lambda_y\sum_i(d^\dag_{i+e_y}d^\dag_i-d_{i+e_y}d_i)\nonumber
\end{eqnarray}
where $\tilde J_{x,y}=\frac{J_{x,y}-\lambda_{x,y}}2$ and $\tilde
\lambda_{x,y}=\frac{J_{x,y}+\lambda_{x,y}}2$. The spinless fermion
$d_i=\frac{1}2(\psi_{i_w}+i\psi_{i_b})$ is located at the $z$-link
and all $z$-links form a square lattice. Taking $\lambda_x+J_x=0$
and $\tilde\lambda_y=\kappa$, this Hamiltonian describes a
$p_x+ip_y$-wave pairing state in this square lattice. As we have
shown, the phase diagram consists of two phases: the topologically
trial A phase and non-trivial B phase. The phase boundary is the
lines:$J_z\pm \tilde J_x\mp \tilde J_y$ if we restrict to $J_z>0$
and $\tilde J_{x,y}>0$. The A phase is an abelian anyon phase
which is equivalent to that in the toric code model, which was
studied before \cite{kitaev,cn}. If we do not consider the high
energy Majorana fermion excitation, the effective Hamiltonian of
the A phase in the honeycomb lattice reads \cite{kitaev}
\begin{eqnarray}
H_{eff}=-J_{eff}(\sum_PW_P+\sum_{z-links} A_j), \label{eff}
\end{eqnarray}
where $A_j=\psi_1\psi_3\psi_6\psi_4$ (see Fig. 1) with
$\lambda_{x,y}=\kappa=0$ and $J_{eff}= \frac{J_x^2J_y^2}{16J_z^3}$
for $J_{x,y}\ll J_z$. One may directly check that the ground state
is given by
\begin{eqnarray}
|G_A\rangle\propto \prod_j(1+A_j)\prod_P(1+W_P)|\phi\rangle,
\end{eqnarray}
because $W_P |G_A\rangle=A_j|G_A\rangle=|G_A\rangle$ with
$|\phi\rangle=|\uparrow\cdots\uparrow\rangle$ a reference state.
Here the Majorana fermion excitations have a high energy $2J_z$
and has been neglected. This Hamiltonian has a $Z_2\times Z_2$
gauge symmetry generated by $W_P$ and $A_j$ with $A_j^2=W_P^2=1$
and $[A_j,W_P]=0$. We know that $A_j$ and $W_P$ are corresponding
to the 'electric charge' and 'magnetic charge' in the toric code
model. The low energy excitations have $A_j=-1$ or $W_P=-1$, which
are $e$ and $m$ vortices in the toric code model and  obey the
mutual semion statistics \cite{kitaev}. It was noted that the
fermion excitations may not be ignored in some braiding processes
\cite{vidal} and there is a controversy to this matter recently
\cite{con}.

In the B phase, we do not have a conserved 'electric charge'
$A_j$. Since the ground state sector in the continuous limit is a
$p_x+ip_y$-wave BCS theory, the ground state in the square lattice
may be written down, which is \cite{yu}
\begin{eqnarray}
|G_B\rangle=(\sum_{ i\ne
i'}\frac{1}{z_j-z_{j'}}d^\dag_{j}d^\dag_{j'})^{N/2}={\rm
Pf}(\frac{1}{z_j-z_{j'}})\prod_l d^\dag_{l}|0\rangle\label{pf}
\end{eqnarray}
where $z_j=j_x+ij_y$  is a complex number with $j=(j_x,j_y)$ the
lattice site label. The vacuum state $|0\rangle$ is defined by
$d_j|0\rangle=0$ while the state $|D\rangle=\prod_i
d^\dag_i|0\rangle$ satisfies $d^\dag_j|D\rangle=0$ because the
square lattice is filled. The vacuum state has been written back
in terms of Pauli matrices \cite{cn}, which is given by $
|0\rangle\propto \prod_j(1-Q_j)\prod_P(1+W_P)|\phi\rangle $ due to
$d_j(1-Q_j)=0$ where $Q_j=i\psi_{j_w}\psi_{j_b}$. $1+W_P$ factor
is introduced because $W_P(1+W_P)=1+W_P$ ensures the vortex free
of the ground state, i.e., $W_P|G_B\rangle=|G_B\rangle$.
 Because of $d_j^\dag(1+Q_j)=0$, $d^\dag_j|D\rangle=0$,
which is equivalent to $Q_j|D\rangle=|D\rangle$. Therefore,
\begin{eqnarray}
|D\rangle\propto \prod_j(1+Q_j)\prod_P(1+W_P)|\phi\rangle.
\end{eqnarray}

\noindent{\it Excitations}  A Majorana fermion excitation on the
ground state is given by $\psi_{j_s}|G_{A,B}\rangle$. Due to
$[W_P,\psi_{j_s}]=0$, $\psi_{j_s}|G_{A,B}\rangle$ is also a vortex
free state. According to the original Hamiltonian, the energy cost
to excite a Majorana fermion is given by
\begin{eqnarray}
E_\psi-E_g=\langle
G_{A,B}|\psi_{j_s}|H|\psi_{j_s}|G_{A,B}\rangle-E_g=2J_z,
\end{eqnarray}
where $E_g$ is the ground state energy. Note that $\psi_{i_b}$
relates to $\psi_{i_w}$ by
$\psi_{i_b}|G_{B}\rangle=-i\psi_{i_w}|G_{B}\rangle$.

We are now going to create the vortex excitations which are
defined by $W_P\sigma_P|G_{A,B}\rangle=-\sigma_P|G_{A,B}\rangle$
and $W_{P'}\sigma_P|G_{A,B}\rangle=\sigma_P|G_{A,B}\rangle$ for
$P'\ne P$. Two operators obey these requirements:
\begin{eqnarray}
\sigma^{(1)}_P=\sigma_{i_b}^z\sigma^z_{i_b-2}\sigma^z_{i_b-4}\cdots
,~~\sigma^{(2)}_P=\sigma^y_{i_b}
\sigma^z_{i_b-1}\sigma^z_{i_b-3}\cdots.
\end{eqnarray}
The vortex is located at the plaquette $P$ with $i_b$ being its
'$a_1$' (See Fig. 1). The sites $i_b-1, i_b-2,\cdots$ are also
marked in Fig. 1. One may also define a vortex at the same
plaquette through
$\tilde\sigma^{(1)}_P=\sigma^z_{i_w}\sigma^z_{i_w-2}\sigma^z_{i_w-4}\cdots$
and $\tilde\sigma^{(2)}_P=\sigma^x_{i_w}
\sigma^z_{i_w-1}\sigma^z_{i_w-3}\cdots$.

Creating a single vortex costs an infinite energy, e.g., in the B
phase
\begin{eqnarray}
\langle
G_B|\sigma^{(1)}_P|H|\sigma^{(1)}_P|G_B\rangle=E_g+2(J_x+J_y)(i_b-(-\infty)).
\label{9}
\end{eqnarray}
Therefore, it is impossible to excite a single vortex. However,
exciting a pair of vortices spends finite energy which is
dependent on the difference of the site labels of two vortices,
e.g., two adjacent vortices
$(\sigma_{i_b}^z\sigma^z_{i_b-2}\sigma^z_{i_b-4}\cdots)\cdot
(\sigma_{i_b+2}^z\sigma^z_{i_b}\sigma^z_{i_b-2}\cdots)=\sigma_{i_b+2}^z$,
which costs energy $4(J_x+J_y)$. In the A phase, the energy cost
of a pair of $\sigma^{(1)}$ is $2J_{eff}$, which is much lower
than the energy cost to excite $\psi$ as $J_z\gg J_{x,y}$.
Exciting pairs $\sigma^{(1)}_P\sigma^{(2)}_{P'}$ and
$\sigma^{(2)}_P\sigma^{(2)}_{P'}$ costs energy $2(2J_{eff}+J_z)$
and $4(J_{eff}+J_z)$, which are also the high energy excitations.

\noindent{\it Fusion rules}  To see nonabelain anyonic fusion
rules, we focus on the B phase to study the fusion rules. Since
$\psi_{i_s}^2=1$, the fusion rule of the Majorana fermions is
$\psi\times \psi=1$. For the vortex excitations, one has
\begin{eqnarray}
&&\sigma_P^{(1)}\cdot\sigma_P^{(1)}=1,~~
\sigma_P^{(2)}\cdot\sigma_P^{(2)}=1,~~\sigma_P^{(1)}\cdot\sigma_P^{(2)}\propto
\psi_{i_b},\nonumber\\
&&\psi_{i_b}\cdot\sigma_P^{(1)}\propto \sigma_P^{(2)},~
\psi_{i_b}\cdot\sigma_P^{(2)}\propto \sigma_P^{(1)}. \label{abel}
\end{eqnarray}
 Define two vortex operators $\sigma_P= \alpha\sigma^{(1)}_P+
\beta\sigma^{(2)}_P$ and $\sigma^\dag_P=\alpha^*\sigma^{(1)}_P+
\beta^*\sigma^{(2)}_P$, which obey
$\sigma_P\sigma^\dag_P=|\alpha|^2+|\beta|^2+i(\alpha\beta^*-
\alpha^*\beta)\psi_{i_b}$. We can not distinguish vortex pairs
$\sigma_P\sigma_{P'}$, $\sigma^\dag_P\sigma^\dag_{P'}$,
$\sigma_P\sigma^\dag_{P'}$ and $\sigma^\dag_P\sigma_{P'}$ because
they are energetically degenerate. This means the equivalence
between $\sigma_P$ and $\sigma^\dag_P$ and the fusion rule for the
vortices is $\sigma\times \sigma=1+\psi$. On the other hand, the
paired vortices are described by a Pfaffian wave function
\cite{mr,yu}
\begin{eqnarray}
{\rm
Pf}\biggl[\frac{(z_j-z_{i_P})(z_{j'}-z_{i_{P'}})+(j\leftrightarrow
j')}{z_j-z_{j'}}\biggr]\label{wpf},
\end{eqnarray}
no matter what kind two vortices are. This also implies the
equivalence between $\sigma_P$ and $\sigma^\dag_P$. Summarily, the
fusion rules in the B phase are
\begin{eqnarray}
\psi\times\psi=1,~~\sigma\times \sigma=1+\psi,~~
\psi\times\sigma=\sigma.
\end{eqnarray}
These nonabelian fusion rules are the same as those in the Ising
model. All fusions cost energy in the same order as that to create
a $\psi$ and a pair of vortices.

\noindent{\it Braiding matrix} The Majorana fermions are
anti-commutative which gives $R^{\psi\psi}_1=-1$. To see the
braiding matrix elements for the vortices, we rotate vortices
counterclockwise. In the B phase, we consider two vortices at
$P_2$ and $P_3$ in Fig. 1. We have two ways to rotate them. One
way is moving the vortex at $P_2$ to $P_1$ first, then moving the
vortex at $P_3$ to $P_2$, and moving that at $P_1$ to $P_3$ (See
Fig. 2(up panel)). This exchanges two vortices and the three steps
is given by acting
$\sigma^y_{j_b}\cdot\sigma^z_{j_b}\cdot\sigma^x_{j_b}$ in turn on
$\sigma_{P_2}\sigma_{P_3}|G_B\rangle$. A single $\sigma^{x,y}$
action means, while one vortex is moving, (say
$\sigma_{P_2}\to\sigma_{P_1}$) another one, say $\sigma_{P_3}$,
becomes $\sigma'_{P_3}=\psi\times\sigma_{P_3}$. Thus, this
exchange is accompanied by non-trivial fusions or creation and
annihilation of the Majorana fermions and then the braiding matrix
element is denoted by
$R^{\sigma\sigma}_\psi=\sigma^y_{j_b}\sigma^z_{j_b}\sigma^x_{j_b}=i$.
Rotating vortices clockwise, we have
${R^{\sigma\sigma}_\psi}^{-1}=\sigma^x_{j_b}\sigma^z_{j_b}\sigma^y_{j_b}=-i$.
This also means $ R^{\sigma\psi}_\sigma=R^{\psi\sigma}_\sigma=-i$.
Another way to exchange two vortices are moving the vortex at
$P_2$ to $P_1$ and that at $P_3$ to $P_4$ simultaneously, then
moving that at $P_1$ to $P_3$ and $P_4$ to $P_1$ in the same time
(See Fig. 2(low panel)). This exchange corresponds to
$\sigma^x_{j_w}\sigma^y_{j_b}\sigma^x_{j_b}\sigma^y_{j_w}\sigma_{P_2}\sigma_{P_3}|G_B\rangle
=\sigma^z_{j_w}\sigma^z_{j_b}\sigma_{P_2}\sigma_{P_3}|G_B\rangle
=\sigma_{P_2}\sigma_{P_3}|G_B\rangle$ since
$\sigma^z_{j_b}\sigma^z_{j_w}|G_B\rangle=|G_B\rangle$.
Simultaneous $\sigma^y_{j_w}$ and $\sigma^x_{j_b}$ action creates
two Majorana fermions $\psi_{j_b}$ and $\psi_{j_w}$ but the
relation $i\psi_{j_w}\psi_{j_b}|G_B\rangle=|G_B\rangle$ means no
fermion is created at any stage \cite{vidal}.
 This defines a braiding matrix element $R^{\sigma\sigma}_1=1$
to exchange. Therefore, the nonabelian braiding matrix for the B
phase is given by
\begin{eqnarray}
R^{\psi\psi}_1=-1,~R^{\sigma\sigma}_1=1,~
R^{\sigma\psi}_\sigma=R^{\psi\sigma}_\sigma=-i,
~R^{\sigma\sigma}_\psi=i.~~ \label{braid}
\end{eqnarray}
Here we do not put in the abelian phase factor $e^{-i\pi/8}$ which
comes from the Pfaffian (\ref{wpf}). Restoring this phase factor,
one has $R^{\sigma\sigma}_1=e^{-i\pi/8}$ and
$R^{\sigma\sigma}_\psi=e^{i3\pi/8}$ and the braid matrix
 are the same as that of the Ising
model \cite{ms,w,kitaev}.

\begin{figure}[htb]
\begin{center}
\includegraphics[width=7cm]{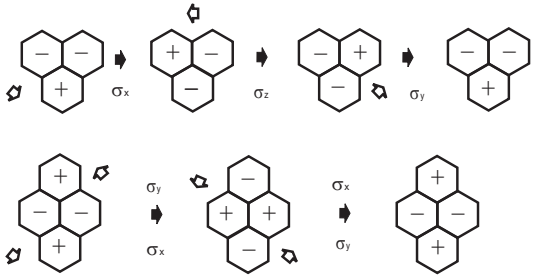}
\end{center}
\vspace{-0.4cm}
 \caption{\label{fig:Fig.2}
Exchange of the vortices.  '$-$' at a plaquette labels a vortex.
Pauli matrix above the filled arrow means the hermitian rotation
operator acting on the white site '5' while below the filled arrow
means the operator acting on the black site '2'. Up panel: The
exchange leads to a phase factor $i$ for the B phase. Low panel: a
bosonic exchange for the B phase. In the A phase, the up panel is
forbidden .} \vspace{-0.4cm}
\end{figure}

\noindent{\it Experimental implications} Recently, there are
several proposals to excite, operate and observe the abelian
anyons in the toric code model. Most of them are based on the
atoms or molecules in optical lattice
\cite{zoller,cirac,bloch,zd,vidal}. Since the nonabelian braiding
matrix (\ref{braid}) is not related to the Pfaffian factors in
eqs. (\ref{pf}) and (\ref{wpf}), all the states involved in do not
have the site-dependent coefficients if we neglect the Pfaffians.
Thus, all techniques applied to the toric code model may be
employed to the present model. For example, load cold atoms in a
honeycomb optical lattice  and manipulate an ancillary atom as
proposed in Ref.\cite{cirac}.

We can also design a possible proof-in-principle experiment to the
nonabelian anyons by using the systems to prove the abelian anyons
in the toric code model \cite{han,pan,pachos,du}.  As Han et al
designed a scheme to demonstrate the abelian anyons in the toric
code model, the minimal lattice needed to rotate or exchange two
vortices are six sites connected, e.g., by the the thick links in
Fig. 1. For this minimal lattice, we prepare the state $|D\rangle$
\begin{eqnarray}
&&|D\rangle\propto
(1+\sigma^y_2\sigma^z_3\sigma^z_4\sigma^y_5)(1+\sigma^z_1\sigma^y_2\sigma^z_4\sigma^x_5)
(1+\sigma^x_2\sigma^y_5\sigma^z_6\sigma^z_3)
\nonumber\\&&(1+\sigma_1^y\sigma^x_3)
(1+\sigma_4^x\sigma^y_6)|\phi\rangle.
\end{eqnarray}
This is an entangled state of 28 pure states and a bit complicated
to be prepared experimentally but is still accessible. The two
vortices state may be given by, e.g., $\sigma_2^z|D\rangle$ which
creates two vortices at $P_2$ and $P_3$ because
$\{\sigma^z_2,\sigma^z_1\sigma^y_2\sigma^z_4\sigma^x_5\}
=\{\sigma^z_2,\sigma^x_2\sigma^y_5\sigma^z_6\sigma^z_3\}=0$ and
$[\sigma^z_2,\sigma^y_1\sigma^z_2\sigma^x_3]=0$. Two exchanges
described before, $\sigma^x_5\sigma^y_2\sigma^x_2\sigma^y_5 $ and
$\sigma^y_2\sigma^z_2\sigma^x_2$, result in the signs $\pm$,
respectively. A full nonabelian two vortices includes a Pfaffian
factor (\ref{wpf}), which contributes an abelian phase factor. We
do not know if it is possible or not to prepare such a state.
Fortunately, to see the nonabelian braiding matrix (\ref{braid}),
one needs to simply act a two vortex operator on $|D\rangle$ and
check those two different exchanges are enough.

\noindent{\it Conclusions } We have converted the ground state and
elementary excitations from the fermionic representation in a
square lattice to the original spin representation in the
honeycomb lattice for the Kitaev-type model. Pauli matrix version
of these states leads to an explicit demonstration to the
non-abelian statistics of anyons in this model. We showed the
nonabelain fusion rules and calculated the non-abelian braiding
matrices. We proposed a proof-in-principle experiment to create,
manipulate and detect the nonabelain anyons in nature.

The authors thank Ville Lahtinen and Julien Vidal for the helpful
comments. This work was supported in part by the national natural
science foundation of China, the national program for basic
research of MOST of China and a fund from CAS.

\end{document}